\documentclass{PoS}

\usepackage[small]{caption}
\usepackage{subcaption}
\usepackage{amsmath, amssymb,graphics,epstopdf,color,cancel,graphicx}
\usepackage[utf8]{inputenc}

\usepackage{lineno}
\title{Searching for TeV DM evidence from Dwarf Irregular Galaxies with the HAWC Observatory}

\ShortTitle{HAWC DM Upper Limits for dIrr Galaxies}

\author{\speaker{Sergio Hern\'andez Cadena}, Rub\'en Alfaro, Andr\'es Sandoval, Ernesto Belmont, Hermes Le\'on\\%\thanks{A footnote may follow.}\\
        Instituto de F\'isica, UNAM, Ciudad de M\'exico, Mex.\\
        E-mail: \email{skerzot@gmail.com, ruben@fisica.unam.mx, asandoval@fisica.unam.mx, belmont@fisica.unam.mx, hleonvar@fisica.unam.mx}}

\author{Viviana Gammaldi$^{a,b}$, Ekaterina Karukes$^{c}$, Paolo Salucci$^{a,b}$\\
        $^a$  Scuola Internazionale Superiore di Studi Avanzati(SISSA), Trieste, Italy\\
        $^b$ Instituto Nazionale di Fisica Nucleare (INFN)\\
        $^c$ ICTP-SAIFR \& IFT-UNESP, R. Dr. Bento Teobaldo Ferraz 271, S\~ao Paulo, Brazil\\
        E-mail: \email{vgammald@sissa.it, ekaterina.karukes@sissa.it, paolo.salucci@sissa.it}}

\author{for the HAWC Collaboration\\
        For a complete author list, see http://www.hawc-observatory.org/collaboration/icrc2017.php\\
        }

\abstract{The dynamics of dwarf irregular (dIrr) galaxies are observed to be dominated by dark matter (DM). Recently, the DM density distribution has been studied for 31 dIrrs. Their extended DM halo (Burket type profile) makes these objects good candidates for DM searches. Located in Puebla (Mexico), the High Altitude Water Cherenkov (HAWC) Observatory is an optimal instrument to perform such DM searches, because of its  large sky coverage (8.4 sr per day). We analyzed a set of two years of HAWC data and we found no significant DM signal from dIrr galaxies. We present the upper limits for DM annihilation cross-section with dIrr galaxies.}

\FullConference{35th International Cosmic Ray Conference -ICRC217-\\
		10-20 July, 2017\\DM
		Bexco, Busan, Korea}

\begin{document}

\section{Introduction}\label{intro}

%Karukes and Salucci (reference), using available data for rotation curves,
It has been shown that local ($d<11$Mpc) Dwarf Irregular (dIrr) Galaxies  are Dark Matter (DM) dominated systems \cite{karukes}. Using available data for rotation curves, the DM distribution is constrained to be described by an spherical DM halo with a Burkert density profile which extends up to radius greater than 15 times the optical radius $r_{opt}$\footnote{The optical radius is the radius where the baryonic or luminous matter is contained}. These Galaxies can be treated  as systems with null gamma-ray emission because dIrr galaxies have a very low star-formation rate ($\thicksim0.003M_{\odot}yr^{-1}$) \cite{thesis}, \cite{stellar_pop}, low metalicity and are composed of stars with masses between 1 and 3 $M_{\odot}$. Although it has been proposed that Stellar Formation Regions can present gamma-ray emission with energies of several hundreds of GeV up to several TeV, the necessary conditions for this processes require that the young stars must be super-massive ($M>8M_{\odot}$) and/or have very strong stellar winds that can produce collective strong chock bubbles where charged particles can efficiently accelerated up to relativistic velocities \cite{hess}. These conditions are not fulfilled by dIrr Galaxies, so any detected gamma-ray emission from these sources could be an indication of the annihilation of DM particles in the halo. %In summary, at TeV energies, dIrr Galaxies are similar to dSph Galaxies and they are a good targets to perform DM searches.
%Because there are no reported Very High Energy (VHE) observations dIrr Galaxiesand  are good candidates to perform indirect DM searches and constrain the value of the cross-section for annihilation of WIMPs from a novel source population. 
In this sense, dIrr galaxies  can be analyzed in a similar way to dSph galaxies and are another way to look for DM. In this work, we show for the first time the upper limits for annihilation cross-section of WIMPs with masses between 1 TeV and 100 TeV for 31 dIrr Galaxies within the HAWC field-of-view. The data set comprises about 2 years of data. 
%field-of-view and with 2 years of data of the HAWC Observatory.

\section{The HAWC Observatory}\label{hawc}

Located in Sierra Negra, Mexico at an altitude of 4100 meters, the High Altitude Water Cherenkov (HAWC) Observatory is an extended array of 300 Water Cherenkov Detectors (WCD) to detect air showers produced by VHE gamma-rays. Every WCD is 7.3m in diameter and 4.5m deep, filled with 200,000L of purified water and is instrumented with 4 Photo-Multiplier Tubes (PMT) to collect the Cherenkov light produced by charged particles passing through the WCDs. The HAWC Observatory has an instantaneous field of view of $\thicksim$2 sr and a duty cycle $>95\%$, and is sensitive to gamma-rays with energies in the range from 1 TeV to 100 TeV. Therefore it is able to investigate the emission from several kind of sources \cite{hawc_sensi, 2hwc-catalogue}, such as AGNs \cite{daily_monitoring}, GRBs \cite{hawc_grb1, hawc_grb2}, PWNs \cite{hawc_crab}, and, in particular, the expected production of photons from annihilating or decaying DM \cite{hawc_dm, hawc_dsph}.

\section{DM photon flux}\label{flux}

To compute the expected gamma-ray flux from DM annihilation, the information about the structural properties of the DM halo and the production mechanisms of photons are needed. The differential gamma-ray flux produced from a source is:
%in a  volume $V$ is:

\begin{equation}
\frac{\mathrm{d}\Phi_{(\gamma)}}{\mathrm{d}E}\,=\,\underbrace{\frac{\langle\sigma v\rangle}{8\pi m_{\chi}^2}\,\frac{\mathrm{d}N_{(\gamma)}}{\mathrm{d}E}}_{\text{Particle Physics Factor}}\,\underbrace{\int_{\Delta\Omega}\int_{l.o.s.}\,\mathrm{d}l\mathrm{d}\Omega\,\rho\left(r(l)\right)^{2}}_{\text{Astrphysical Factor $\mathrm{J}$}}
\label{dm_flux}
\end{equation}

\hspace{-0.75cm}where $\mathrm{d}N_{(\gamma)}/\mathrm{d}E$ is the differential spectrum of photons produced for an annihilation channel, $m_{\chi}$ is the mass of the WIMP, $\langle\sigma v\rangle$ is the thermal averaged annihilation cross-section, and $\rho\left(r(l)\right)$ is the  DM density profile. The second term in equation \ref{dm_flux} is the Astrophyical Factor $\mathrm{J}$, or $\mathrm{J}$ factor. The $\mathrm{J}$ factor is defined as the integral of the DM density ($\rho$) squared along the line of sight ($l.o.s.$) distance $l$ and over the solid angle around the line of sight. For all the calculations, we assumed that the dIrr Galaxies are small and there is not appreciable contribution of sub-structures to $\mathrm{J}$ factors. The $\mathrm{J}$ factors are calculated over all the spatial extension of the sources and are listed in Table \ref{dIrr_galaxies}.

\subsection{Photon Spectra}\label{spectre}

\begin{figure}[t!]
\captionsetup{width=0.8\linewidth}
\centering
\includegraphics[width=0.9\linewidth]{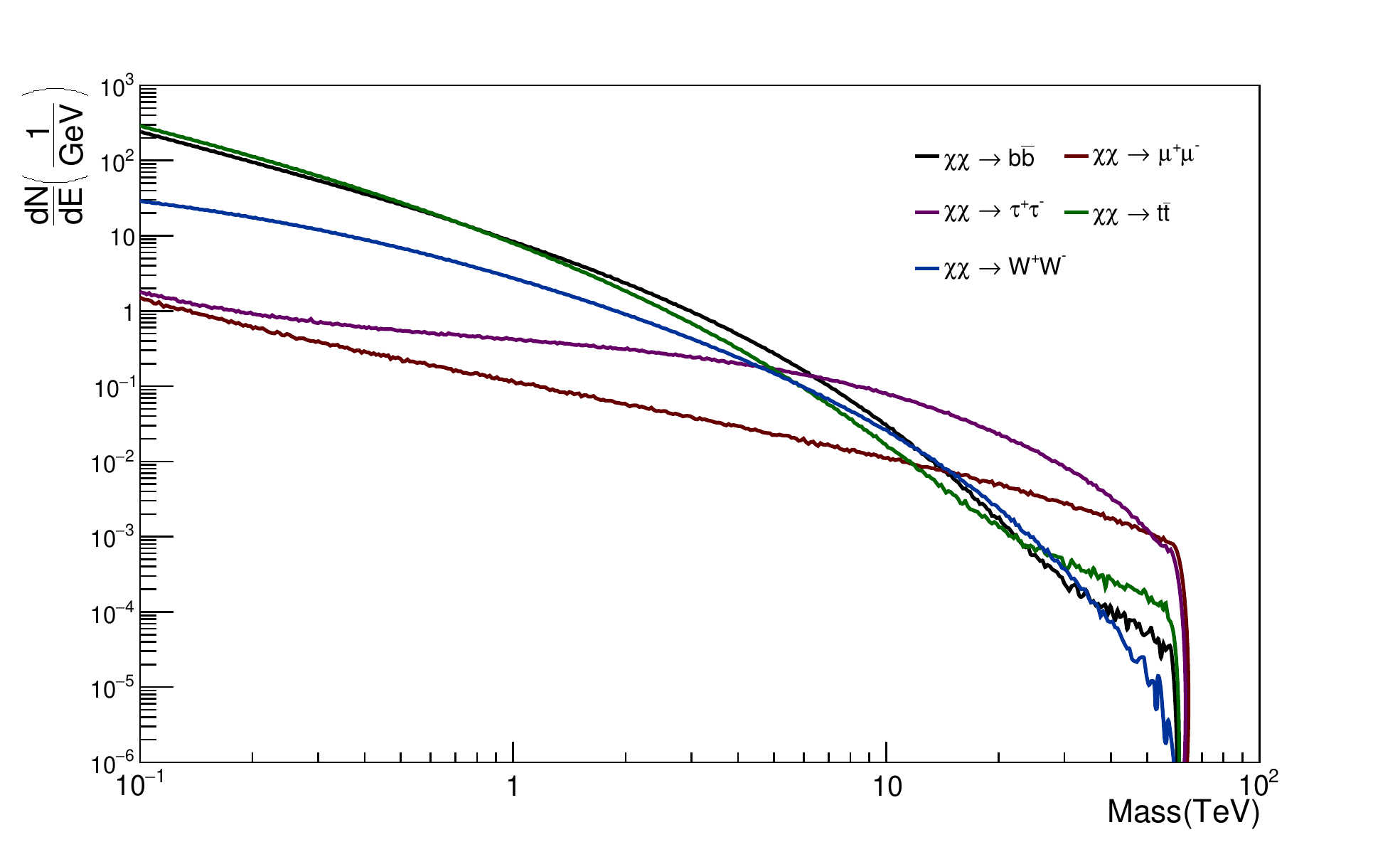}
\caption{Spectrum of Photons computed for annihilation of WIMPs with mass of 60 TeV to five channels, assuming that the branching ratio for every channel is 100\%.}
\label{spectra}
\end{figure}

The production of photons from annihilation of DM particles is due to the decay or hadronization processes of the unstable products. The spectrum is continuous and has an energy cut-off at the mass of the DM particle. For this work, we considered WIMP masses in the range from 1 TeV to 100 TeV and annihilation to five channels: $b$ and $t$ quarks, $\mu$ and $\tau$ leptons, and the $W$ boson. The spectrum of photons is calculated with \textsc{Pythia 8} \cite{pythia}. In Figure \ref{spectra} the spectrum of photons for a WIMP with mass $m_{\chi}=60$ TeV is shown.

\begin{table}[t!]
\captionsetup{width=0.9\linewidth}
\centering
\begin{tabular}{c||c|c|c|c|c|c|c|c}
Name & $\alpha$  & $\delta$ & $\frac{M_{\text{DM}}}{M_{\text{Lum}}}$ & $R_{vir}$  & $\rho_0$ & $r_0$ & $\mathrm{J}$ & $\sigma$\\
& $(^{\circ})$ & $(^{\circ})$ & & (kpc) & $M_{\odot}/\text{kpc}^3$ & (kpc) & $\text{TeV}^2/\text{cm}^5$\\
\hline
\hline
UGC1281 & 27.3833 &32.5925 & 66.80 & 82.1 & 4.04$\times10^{7}$ & 2.93 & 9.2572$\times10^9$ & $-1.1490$ \\
UGC1501 & 30.3167 & 28.8436 & 72.30 & 89.5 & 1.76$\times10^{7}$ & 4.32 & 6.1530$\times10^9$ & $-1.3726$\\
UGC5427 & 151.1708 &29.3664 & 48.21 & 54.1 & 46.7$\times10^{7}$ & 0.76 &1.0986$\times10^{10}$ & $+0.6616$\\
UGC7559 & 186.7708 & 37.1425 & 57.51 & 59.6 & 2.67$\times10^{7}$ & 2.45 & 2.4702$\times10^9 $ & $-0.8200$\\
UGC8837 & 208.6875 & 53.9047 & 75.33 & 92.7 & 1.09$\times10^{7}$ & 5.40 & 1.9164$\times10^9$ & $-1.2860$\\
UGC7047 & 181.0083 & 52.5886 & 50.03 & 48.8 & 7.70$\times10^{7}$ & 1.34 & 4.4390$\times10^9$ & $-1.7911$\\
UGC5272 & 147.5917 & 31.4875 & 72.65 & 94.9 & 2.30$\times10^{7}$ & 4.14 & 4.1065$\times10^9$ & $-1.0884$\\
DDO52  & 127.1167 & 41.8567 & 76.17 & 102.3 & 2.63$\times10^{7}$ & 4.24 & 2.8218$\times10^9$ & $+0.3556$\\
DDO101 & 177.9125 & 31.5194 & 67.15 & 86.8 & 5.52$\times10^{7}$ & 2.71 & 1.2689$\times10^9$ & $-1.0574$\\
DDO154 & 193.520 & 27.1486 & 54.64 & 56.3 & 4.02$\times10^{7}$ & 1.60 & 4.3494$\times10^9$ & $+0.2617$\\
DDO168 & 198.6167 & 45.9194 & 64.48 & 83.3 & 7.78$\times10^{7}$ & 2.29 & 2.1513$\times10^{10}$ & $-0.4791$\\
Haro29 & 186.5667 & 48.4919 & 40.08 & 33.0 & 35.8$\times10^{7}$ & 0.51 & 3.0811$\times10^9$ & $+0.7125$\\
Haro36  & 191.7333 & 51.6131 & 67.11 & 85.5 & 4.71$\times10^{7}$ & 2.84 & 3.6515$\times10^9$ & $-1.0719$\\
IC10 & 5.100 & 59.2917 & 45.74 & 44.5 & 25.8$\times10^{7}$ & 0.78 & 3.3740$\times10^{11}$ & $-1.3696$\\
WLM & 0.4917 & -15.4611 & 48.12 & 44.2 & 6.58$\times10^{7}$ & 1.29 & 4.3553$\times10^{10}$ & $-1.2415$\\
UGC7603 & 187.1833 & 22.8225 & 71.84 & 95.6 & 3.86$\times10^{7}$ & 3.42 & 4.7883$\times10^9$ & $-0.4815$\\
UGC7861 & 190.4667 & 41.2739 & 58.29 & 73.8 & 16.7$\times10^{7}$ & 1.52 & 9.0994$\times10^9$ & $-1.5236$\\
DDO125 & 186.9208 & 43.4939 & 38.75 & 25.5 & 2.33$\times10^{7}$ & 1.10 & 5.5139$\times10^8$ & $+0.1665$\\
UGC7866 & 190.5625 & 38.5019 & 46.22 & 39.6 & 5.12$\times10^{7}$ & 1.27 & 1.4874$\times10^9$ & $+0.6805$\\
DDO43 & 112.0708 & 40.7703 & 48.76 & 47.2 & 6.92$\times10^{7}$ & 1.35 & 2.0781$\times10^9$ & $-0.8621$\\
IC1613 & 16.1958 & 2.1333 & 41.92 & 30.1 & 1.75$\times10^{7}$ & 1.46 & 5.8729$\times10^9$ & $-0.6184$\\
NGC6822 & 296.2375 & -14.8031 & 49.17 & 46.6 & 7.09$\times10^{7}$ & 1.32 & 1.1446$\times10^{11}$ & $-0.5364$\\
UGC7916 & 191.1042 & 34.3864 & 69.89 & 77.7 & 0.58$\times10^{7}$ & 7.52 & 4.3566$\times10^8$ & $-1.2203$\\
UGC5918 & 162.4000 & 65.5306 & 67.98 & 79.6 & 1.73$\times10^{7}$ & 3.88 & 1.7454$\times10^9$ & $-0.3268$\\
AndIV & 10.6250 & 40.5758 & 45.90 & 40.8 & 9.07$\times10^{7}$ & 0.99 & 1.0871$\times10^9$ & $+0.0012$\\
UGC7232 & 183.4333 &36.6333 & 37.67 & 32.2 & 96.4$\times10^{7}$ & 0.34 & 2.8365$\times10^{10}$ & $+1.2456$\\
DDO133 & 188.2208 & 31.5392 & 60.22 & 66.4 & 3.20$\times10^{7}$ & 2.55 & 4.0290$\times10^9$ & $+0.9029$\\
UGC8508 & 202.6850 & 54.9100 & 37.09 & 27.4 & 22.6$\times10^{7}$ & 0.50 & 5.2128$\times10^9$ & $+0.8416$\\
UGC2455 & 194.9267 & 25.2375 & 65.77 & 77.4 & 2.62$\times10^{7}$ & 3.21 & 2.1165$\times10^9$ & $-0.2209$\\
NGC3741 & 174.0267 & 45.2853 & 33.53 & 21.4 & 65.7$\times10^{7}$ & 0.26 & 3.8677$\times10^9$ & $+0.0136$\\
UGC11583 & 307.5637 & 60.4402 & 70.26 & 87.7 & 2.55$\times10^{7}$ & 3.67 & 5.0994$\times10^9$ & $-1.6305$\\
\end{tabular}
\caption{Astrophysical parameters and structural properties for the 31 dIrr Galaxies within the field-of-view of the HAWC Observatory. The source, right ascension ($\alpha$), declination ($\delta$), DM mass to luminous mass ratio, virial radius ($R_{vir}$), scale density $\rho_0$, scale radius $r_0$, the Astrophysical Factor $\mathrm{J}$ are listed above. The quantities related to the DM halo are taken from \cite{karukes} and are computed assuming that density profile is described by a Burkert profile. The $\mathrm{J}$ factors are calculated over all the spatial extension of each source. All the significances $\sigma$ (last column) are obtained for a WIMP with mass $m_{\chi}=60TeV$ and the $\chi\chi\,\to\,\tau^{+}\tau^{-}$ annihilation channel.}
\label{dIrr_galaxies}
\end{table}

\subsection{The DM profile in dIrr Galaxies}

%That dIrr Galaxies are DM dominated systems is not a new idea. 
With the proof that Large and Small Magellanic Clouds  are rotating systems \cite{kerr_1, kerr_2}, measurements of the rotation curve of other dIrr Galaxies were performed. It was observed that the rotation curve for these galaxies flatten after about 2 disk scale lengths and no cases of declining curves were found, indicating that dIrr Galaxies have no bulge. Early results indicated that they are objects dominated by DM at all radii and properties of their DM halos seem to be similar to those from galaxy halos \cite{carignan}.

For the set of dIrr Galaxies we studied here, the structural properties of luminous and DM contributions are constrained using kinematical data taken from \cite{karukes}. The DM density is constrained to be described by a Burkert Profile  \cite{burkert}. The Burkert profile is a density distribution that resembles an isothermal profile in the inner regions ($r<r_0$) and a distribution with slope -3 in the outer regions:

\begin{equation}
\rho(r)\,=\,\frac{r_0^3\rho_0}{(r+r_0)(r^2+r_0^2)}
\label{burkert}
\end{equation}

\hspace{-0.75cm}where $r_0$ and $\rho_0$ are the scale radius and density. Unlike Navarro-Frenk-White (NFW) density profile, the Burkert profile is a cored profile. The coordinates and DM parameters ($r_0,\,\rho_0\,\text{and}\,R_{vir}$) for the 31 dIrr Galaxies are listed in Table \ref{dIrr_galaxies}. The virial radius $R_{vir}$ is computed assuming an overdensity parameter $\Delta=100$. Additionally, we computed the DM mass to luminous mass ratio ($M_{\text{DM}}/M_{\text{Lum}}$, column 4 in Table \ref{dIrr_galaxies}) using the masses obtained for fit results in \cite{karukes}.  From these values, it is observed that the mean amount of DM in these galaxies is 56.43 times their luminous matter.

\section{DM Upper Limits on annihilation Cross Section}\label{results}

\begin{figure}[t!]
\captionsetup{width=0.7\linewidth}
\centering
\begin{subfigure}{0.495\linewidth}
\includegraphics[width=1.0\linewidth]{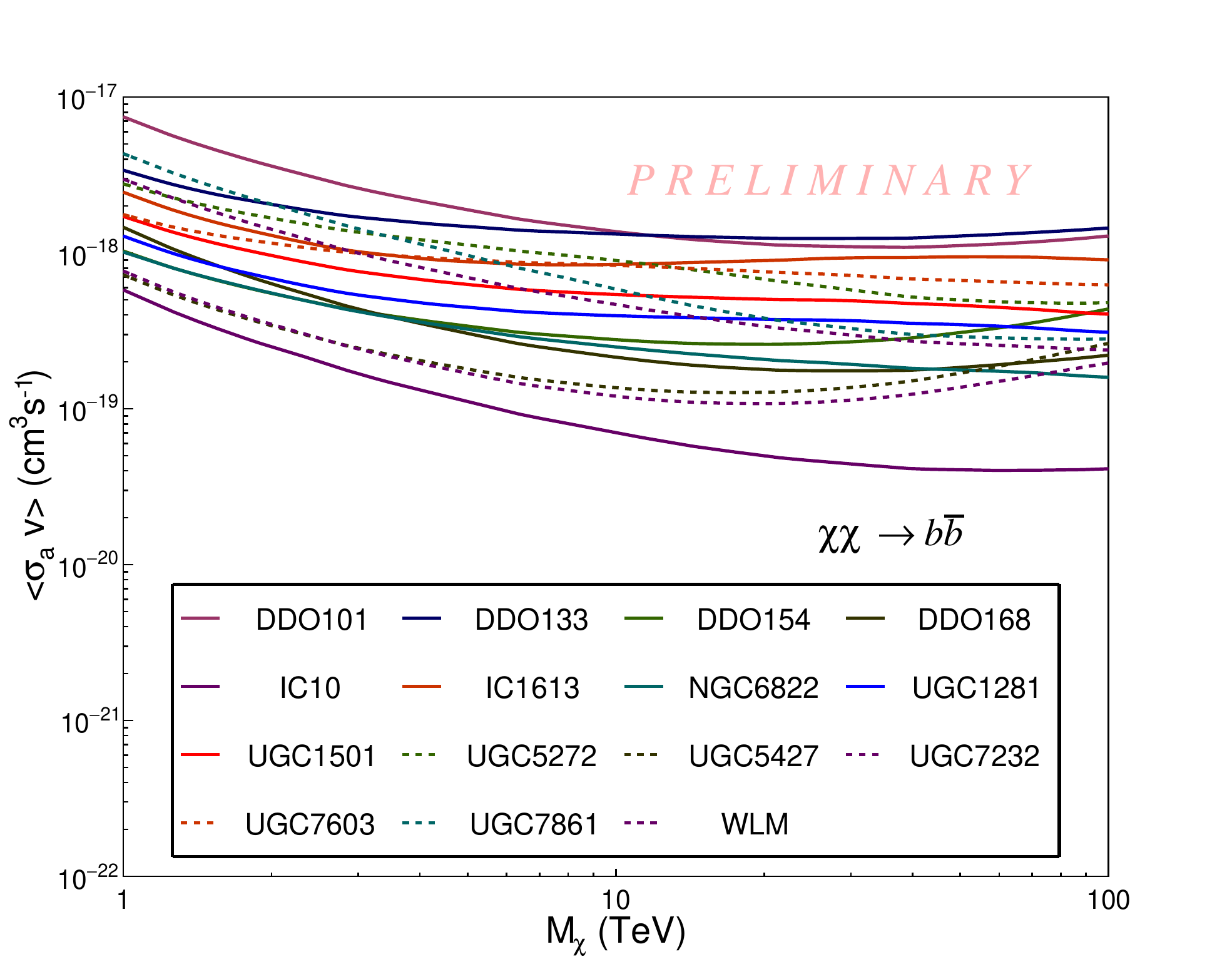}
\end{subfigure}
\begin{subfigure}{0.495\linewidth}
\includegraphics[width=1.0\linewidth]{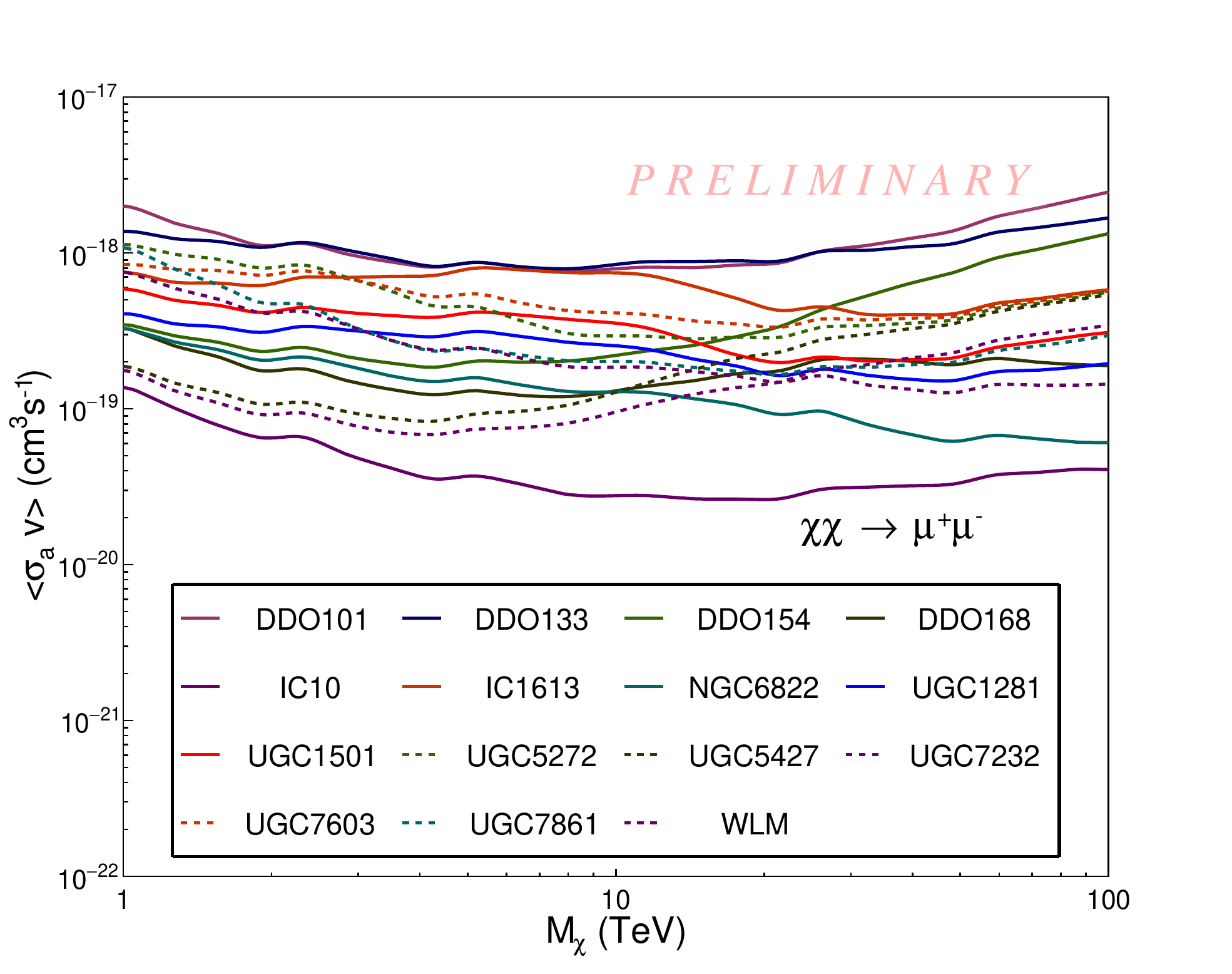}
\end{subfigure}
\vspace{0.35cm}
\begin{subfigure}{0.495\linewidth}
\includegraphics[width=1.0\linewidth]{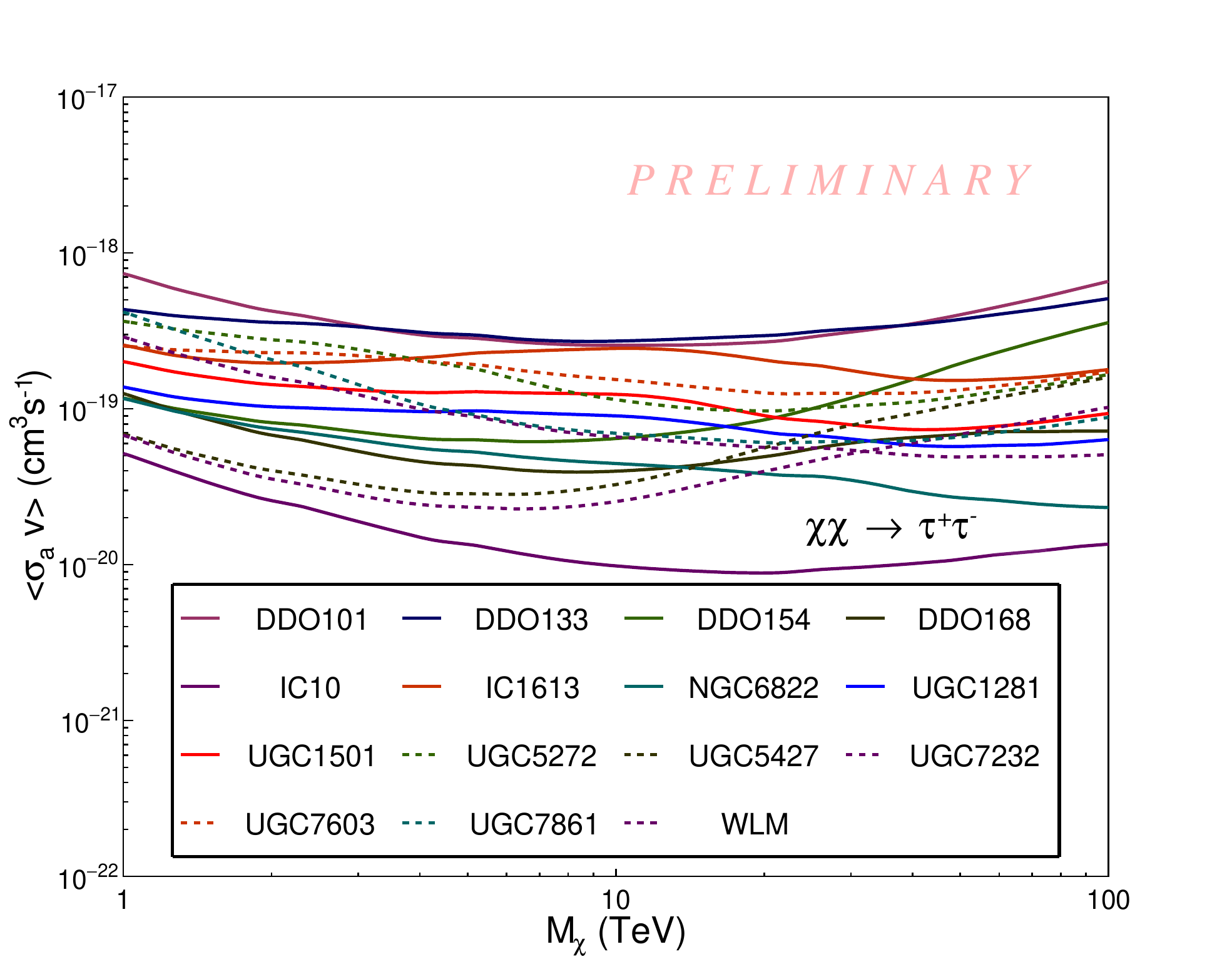}
\end{subfigure}
\begin{subfigure}{0.495\linewidth}
\includegraphics[width=1.0\linewidth]{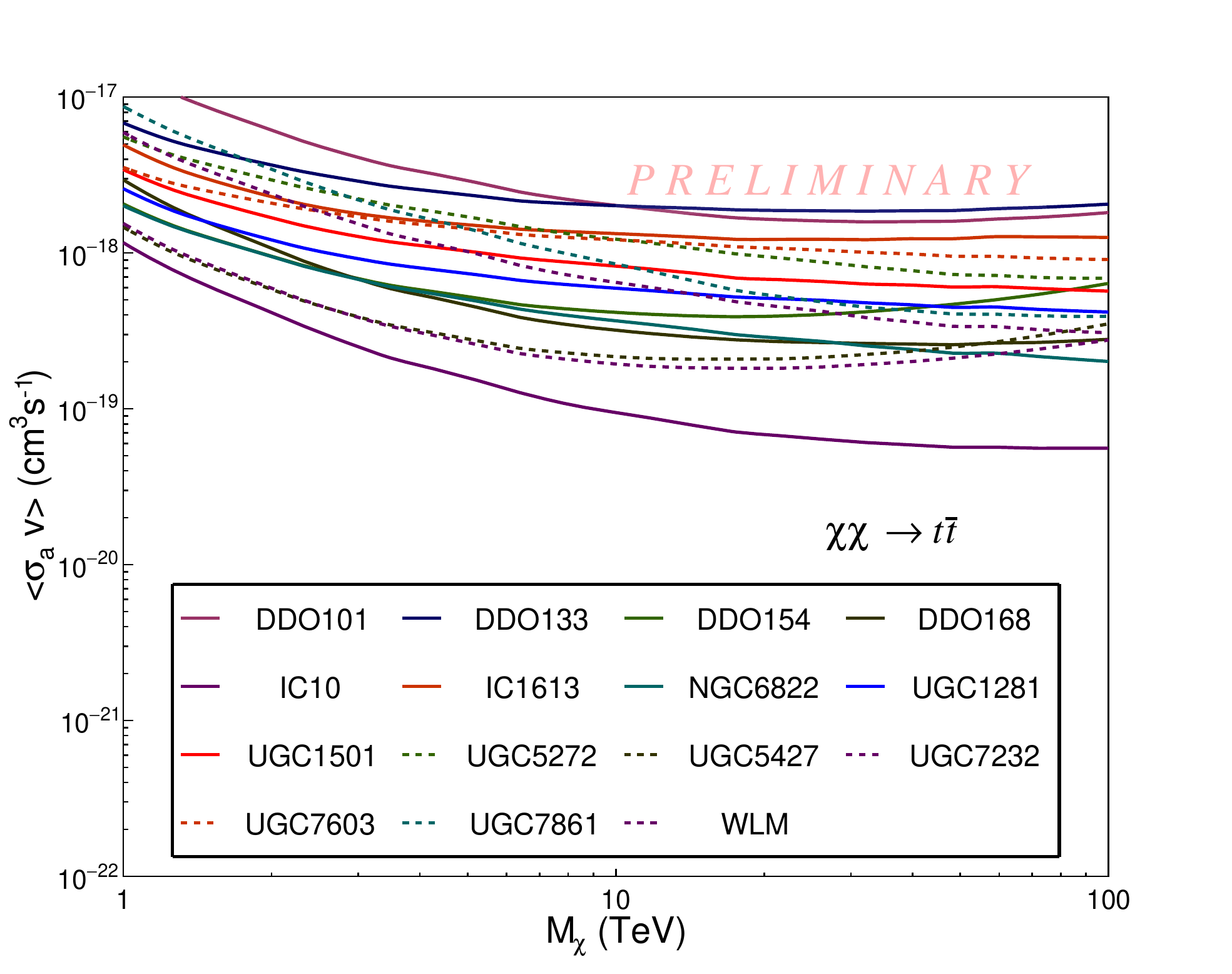}
\end{subfigure}
\vspace{0.35cm}
\begin{subfigure}{0.495\linewidth}
\includegraphics[width=1.0\linewidth]{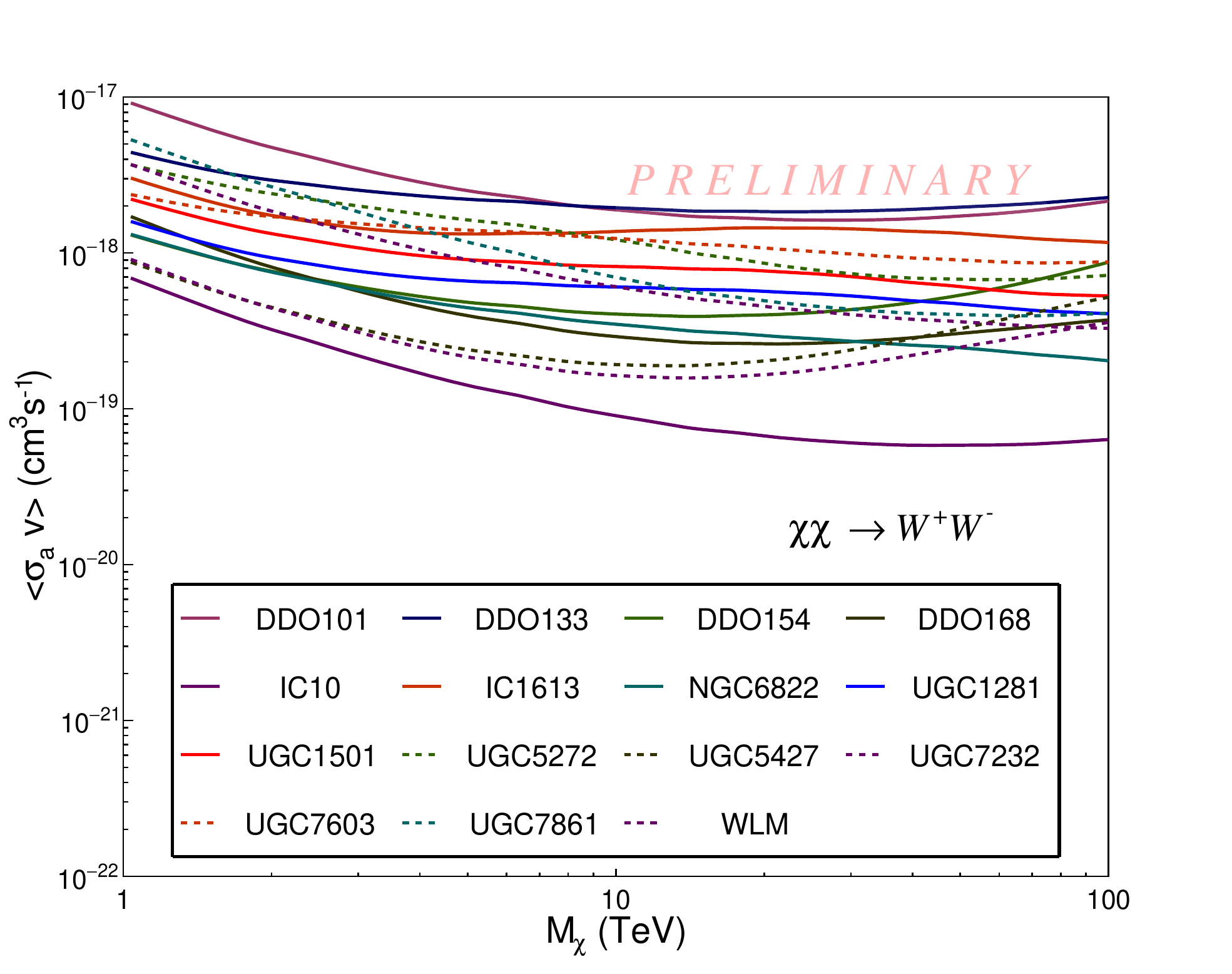}
\end{subfigure}
\caption{Individual upper limits for DM annihilation cross section  for the fifteen best dIrr Galaxies in the field of view of the HAWC Observatory: DDO101, DDO133, DDO154, DDO168, IC10, IC1613, NGC6822, UGC1281, UGC1501, UGC5272, UGC5427, UGC7232, UGC7603, UGC7861, WLM. The region above the curves is excluded.}
\label{upper_limits}
\end{figure}

We calculated the individual upper limits on annihilation cross-section for 31 dIrr Galaxies (see Table \ref{dIrr_galaxies}) within the field-of-view of the HAWC Observatory for 760 days of HAWC data. All the Galaxies were treated as point sources and with no non-DM gamma-ray backgrounds  systems (see section \ref{intro}). %The last assumption is plausible because dIrr Galaxies have a very low star-formation rate ($\thicksim0.003M_{\odot}yr^{-1}$) \cite{thesis}, \cite{stellar_pop}, low metalicity and are composed of stars with masses between 1 and 3 $M_{\odot}$. Although it has been proposed that Stellar Formation Regions can present gamma-ray emission with energies of several hundreds of GeV up to several TeV, the necessary conditions for this processes require that the young stars must be super-massive ($M>8M_{\odot}$) and/or have very strong stellar winds that can produce collective strong chock bubbles where charged particles can efficiently accelerated up to relativistic velocities \cite{hess}. This conditions are not fulfilled by dIrr Galaxies, so any detected gamma-ray emission from these sources could be an indication of the annihilation of DM particles in the halo. In summary, at TeV energies, dIrr Galaxies are similar to dSph Galaxies and they are a good targets to perform DM searches.
Through detailed simulation of the gamma-ray sensitivity and background for the HAWC Observatory, the significance of the gamma-ray flux for a range of DM masses, 1 TeV - 100 TeV, and five annihilation channels is computed. In Table \ref{dIrr_galaxies} we report the significance obtained for the annihilation channel to $\tau$ leptons of a WIMP with mass $m_{\chi}=60$ TeV. Because the HAWC Observatory has not seen statistical significant excess in these regions, the significance for every source is converted into exclusion curves of the annihilation cross-section of WIMPs for the 31 dIrr Galaxies. To get the exclusion curves, we used the Maximum Likelihood Method (for a detailed description of the method, please see \cite{hawc_dsph}) that is implemented in the HAWC software utility \textsc{Liff} \cite{liff}. The upper limits for the fifteen most sensitive dIrr Galaxies are shown in Figure \ref{upper_limits}.

\section{Discussion}\label{discu}

It can be observed from Figure \ref{upper_limits}, that the strongest limits from dIrr Galaxies come from the IC10 Galaxy because its large astrophysical factor $\mathrm{J}$ (3.3740$\times10^{11}\text{TeV}^2\text{cm}^{-5}$). The limits from dIrr Galaxies are comparable to those obtained by HAWC for the Leo I, Leo IV, Hercules and Canes Venatici I Spheroidal Galaxies \cite{hawc_dsph}, and less restrictive than, for example, the limits from Triangulum II and Segue I, whose limits are 4 orders of magnitude more restrictive than the value for IC10, our best target in this sample. This is a consequence of: a) their large astrophysical factors (100-1000 times larger) and b) the suboptimal declination of IC10 (peaking at 40$\deg$ off zenith for HAWC). As it was mentioned in Section \ref{spectre}, $\mathrm{J}$ factors are calculated with a Burkert profile, because this profile reproduces the rotation curves for dIrr Galaxies \cite{karukes}. However, it is possible that cusped profiles can properly  fit the rotation curves by considering non-rotational motions in Galaxies \cite{oman}. We are investigating the effect of considering cusped profiles, such as Einasto or Navarro-Frenk-White \cite{nfw1,nfw2}, in the calculations of $\mathrm{J}$ factors. Future work includes the calculation of lower limits for decay-lifetime $\tau_{\chi}$ of WIMPs for dIrr Galaxies. Because astrophysical factor $\mathrm{D}$ depends only in the amount of DM in the halo, and is less sensitive to profile shape, calculations for dIrr Galaxies, plus a combined analysis, can improve our limits for $\tau_{\chi}$. As an example, the astrophysical factor $\mathrm{D}$ for IC10 is $4.1631\times10^{15}\text{TeV cm}^{-2}$, comparable to the value for Triangulum II.

\section{Summary}\label{sum}

In this work we show the individual 95\% CL limits on the annihilation cross-section for 31 dIrr Galaxies. The limits are calculated for a DM range of masses, 1 TeV -- 100 TeV, and for several DM annihilation channels, $\{b,\mu,\tau,t,W\}$, resulting from data collected over 760 day period. These curves are the first experimental limits calculated for DM searches in dIrr Galaxies in VHE regime. Future work will include a source stacked analysis and an extended source analysis for dIrr Galaxies considered in these proceedings. 

\section{Acknowlegments}

We acknowledge the support from: the US National Science Foundation (NSF); the US Department of Energy Office of High-Energy Physics; the Laboratory Directed Research and Development (LDRD) program of Los Alamos National Laboratory; Consejo Nacional de Ciencia y Tecnolog\'{\i}a (CONACyT), M{\'e}xico (grants 271051, 232656, 260378, 179588, 239762, 254964, 271737, 258865, 243290, 132197), Laboratorio Nacional HAWC de rayos gamma; L'OREAL Fellowship for Women in Science 2014; Red HAWC, M{\'e}xico; DGAPA-UNAM (grants RG100414, IN111315, IN111716-3, IA102715, 109916, IA102917); VIEP-BUAP; PIFI 2012, 2013, PROFOCIE 2014, 2015; the University of Wisconsin Alumni Research Foundation; the Institute of Geophysics, Planetary Physics, and Signatures at Los Alamos National Laboratory; Polish Science Centre grant DEC-2014/13/B/ST9/945; Coordinaci{\'o}n de la Investigaci{\'o}n Cient\'{\i}fica de la Universidad Michoacana. Thanks to Luciano D\'{\i}az and Eduardo Murrieta for technical support. Viviana Gammaldi work has been supported by QGSKY, by the Agencia Estatal de Investigaci\'on (AEI) and Fondo Europeo de Desarrollo Regional (FE-DER) FIS2016-78859-P (AEI/FEFER, UE), by the MINECO (Spain) project FIS2014-52837-P and Consolider-Ingenio MULTIDARK CSD2009-00064, and partially by the H2020CSA Twinning project No. 692194 ORBI-T-WINNINGO. Ekaterina Karukes work was supported by the S\~ao Paulo Research Foundation (FAPESP) under the grant $\#2016/262889.$


\begin{thebibliography}{99}
\bibitem{karukes} E. Karukes \& P. Salucci, \emph{The universal rotation curve of dwarf disc galaxies}, \emph{MNRAS} {\bf{465}}, 4703-4722 (2017) [{\tt{astro-ph.GA/1609.06903v3}}]

\bibitem{thesis} J. M. Dunn, \emph{The Stellar Content and Star-Formation Rates of Dwarf Irregular Galaxies}, 2007

\bibitem{stellar_pop} R. E. Schulte--Ladbeck \& U. Hopp, \emph{The Stellar Content of 10 Dwarf Irregular Galaxies}, \emph{The Astronomical Journal}, {\bf{116}} 2866-2915, 1998 December

\bibitem{hess} S. Ohm, D. Horns, O. Reimer, J. Hinton, G. Rowell, E. de O\~na Wilhelmi, M. V. Fernandes, F. Acero, A. Marcowith for the H.E.S.S. Collaboration \emph{ H.E.S.S. observations of massive stellar clusters} [{\tt{astro-ph.HE/2637v20906}}]

\bibitem{hawc_sensi} A.U. Abeysekara  \textit{et. al.} (The HAWC Collaboration), \emph{Sensitivity of the High Altitude Water Cherenkov detector to sources of multi-TeV gamma rays}, \emph{ApJ}, 50--52 (2013), 26--32

\bibitem{2hwc-catalogue} A.U. Abeysekara  \textit{et. al.} (The HAWC Collaboration), \emph{The 2HWC HAWC Observatory Gamma Ray Catalog}, [{\tt{astro-ph.HE/1702.02992v1}}]

\bibitem{daily_monitoring} A.U. Abeysekara  \textit{et. al.} (The HAWC Collaboration), \emph{Daily Monitoring of TeV Gamma-Ray Emission from Mrk 421, Mrk 501, and the Crab Nebula with HAWC}, \emph{ApJ}, {\bf{841}}:100 (13pp), 2017 June 1

\bibitem{hawc_grb1} A.U. Abeysekara  \textit{et. al.} (The HAWC Collaboration), \emph{Search for Gamma-Rays from the Unsually Bright GRB 130427A with the HAWC Gamma-Ray Oobservatory}, \emph{ApJ}, {\bf{800}}:78 (6pp), 2015 February 20 %800:78 (6pp), 2015 February 20

\bibitem{hawc_grb2} A.U. Abeysekara  \textit{et. al.} (The HAWC Collaboration), \emph{Search for Very-High-Eenergy Emission from Gamma-Ray Bursts Using the First 18 Months of Data from the HAWC Gamma-Ray Observatory}, [{\tt{astro-ph.HE/1705.01551v1}}]

\bibitem{hawc_crab} A.U. Abeysekara  \textit{et. al.} (The HAWC Collaboration), \emph{Observation of the Crab Nebula with the HAWC Gamma-Ray Observatory}, [{\tt{astro-ph.HE/1701.01778v1}}]

\bibitem{hawc_dm} A.U. Abeysekara  \textit{et. al.} (The HAWC Collaboration), \emph{Sensitivity of HAWC to high-mass dark matter annihilations},\emph{Phys. Rev. D},{\bf{90}} 122002 (2014)

\bibitem{hawc_dsph} A. Albert  \textit{et. al.} (The HAWC Collaboration), \emph{Dark Matter Limits From Dwarf Spheroidal Galaxies with The HAWC Gamma-Ray Observatory},\emph{ArXiv e-prints}, [{\tt{astro-ph.HE/1706.01277}}]

\bibitem{pythia} T. Sjostrand, S. Mrenna \& P. Skands, \emph{JHEP} {\bf{05}} (2006) 026, \emph{Comput. Phys. Comm.} {\bf{178}} (2008) 852

\bibitem{kerr_1} F. J. Kerr,  J. V. Hindman \& B. J. Robinson, \emph{Observations of the 21 cm Line from the Magellanic Clouds}, \emph{Ann. J. Phys.} {\bf{7}}, 297, 1954

\bibitem{kerr_2}  F. J. Kerr \& G. De Vacouleurs, \emph{Rotation and other motions of the Magellanic Clouds} \emph{Ann. J. Phys.} {\bf{8}}, 508, 1955

\bibitem{carignan} C. Carignan, R. Sancisi, \& T.S. van Albada, \emph{HI and Mass Distribution in the Dwarf ?regular? Galaxy UGC 2259}, \emph{Astron. J.} 95, 37, 1988

\bibitem{burkert} A. Burkert, \emph{The Structure of Dark Matter Halos in Dwarf Galaxies}, \emph{The Astrophysical Journal} {\bf{447}} L25-L28, (1995) July 1

\bibitem{liff} P. W. Younk, R. J. Lauer, G. Vianello, J. P. Harding, H. A. A. Solares, H. Zhou, M. Hui, for the HAWC Co-
laboration, \emph{A high-level analysis framework for HAWC}, \emph{The 34th International Cosmic Ray Conference},
{\bf{ICRC 2015}}, edited by \emph{PoS} (The Hague, The Netherlands, 2015) p. 948

\bibitem{oman} K. A. Oman, A. Marasco, J. F. Navarro, C. S. Frenk, J. Schaye \& A. Ben\'itez-Lambay, \emph{Apparent cores and non-circular motions in the H I discs of simulated galaxies}, \emph{ArXiv e-prints} [{\tt{astro-ph.HE/1706.07478v1}}]

\bibitem{nfw1} J. F. Navarro, C. S. Frenk \& S. D.  M. White, \emph{A Universal Density Profile from Hierarchical Clustering}, \emph{The Astrophysical Journal}, {\bf{490}}, 493-508, 1997

\bibitem{nfw2} J. F. Navarro, A. Ludlow, V. Springel, J. Wang, M. Vogelsberger, S. D. M. White, A. Jenkins, S. C. Frenk, A. Helmi, \emph{The diversity and similarity of simulated cold dark matter haloes}, \emph{Mon. Not. R. Astron. Soc.} {\bf{402}}, 21-34, 2010

%[Navarro, J. F., Frenk, C. S., & White, S. D. M. 1997, The Astrophysical Journal, 490, 493] and [Navarro, J. F., et al. 2009, arXiv:0810.1522v2 [astro-ph]]
\end{thebibliography}
\end{document}